\def \SAIT #1 #2 {{\em Mem.\ Soc.\ Astron.\ It.\/} {\bf #1}, #2}
\def \MESS #1 #2 {{\em The Messenger\/} {\bf #1}, #2}
\def \ASTRNACH #1 #2 {{\em Astron. Nach.\/} {\bf #1}, #2}
\def \AAP #1 #2 {{\em Astron. Astrophys.\/} {\bf #1}, #2}
\def \AAL #1 #2 {{\em Astron. Astrophys. Lett.\/} {\bf #1}, L#2}
\def \AAR #1 #2 {{\em Astron. Astrophys. Rev.\/} {\bf #1}, #2}
\def \AAS #1 #2 {{\em Astron. Astrophys. Suppl. Ser.\/} {\bf #1}, #2}
\def \AJ #1 #2 {{\em Astron. J.\/} {\bf #1}, #2}
\def \ANNREV #1 #2 {{\em Ann. Rev. Astron. Astrophys.\/} {\bf #1}, #2}
\def \APJ #1 #2 {{\em Astrophys. J.\/} {\bf #1}, #2}
\def \APJL #1 #2 {{\em Astrophys. J. Lett.\/} {\bf #1}, L#2}
\def \APJS #1 #2 {{\em Astrophys. J. Suppl.\/} {\bf #1}, #2}
\def \APSS #1 #2 {{\em Astrophys. Space Sci.\/} {\bf #1}, #2}
\def \ASR #1 #2 {{\em Adv. Space Res.\/} {\bf #1}, #2}
\def \BAIC #1 #2 {{\em Bull. Astron. Inst. Czechosl.\/} {\bf #1}, #2}
\def \JSQRT #1 #2 {{\em J. Quant. Spectrosc. Radiat. Transfer\/} {\bf #1}, #2}
\def \MN #1 #2 {{\em Mon. Not. R. Astr. Soc.\/} {\bf #1}, #2}
\def \MEM #1 #2 {{\em Mem. R. Astr. Soc.\/} {\bf #1}, #2}
\def \PLR #1 #2 {{\em Phys. Lett. Rev.\/} {\bf #1}, #2}
\def \PASJ #1 #2 {{\em Publ. Astron. Soc. Japan\/} {\bf #1}, #2}
\def \PASP #1 #2 {{\em Publ. Astr. Soc. Pacific\/} {\bf #1}, #2}
\def \NAT #1 #2 {{\em Nature\/} {\bf #1}, #2}

\documentstyle{memsait}
\input epsf.sty
\begin{opening}
\title{ MEASUREMENTS OF FAINT SUPERNOVAE}

\author{ROBERT A. SCHOMMER, N.B. SUNTZEFF, R.C. SMITH for the High-Z SN Search
Team}
\institute{Cerro Tololo InterAmerican Observatory, Casilla 603, La Serena
Chile}
\date{} 
\end{opening}

\begin{document}

\oddpagefooter{}{}{} 
\evenpagefooter{}{}{} 
\ 
\bigskip

\begin{abstract}

We summarize the current status of cosmological measurements using 
SNe Ia. Searches to an average depth of z$\sim$0.5 have found approximate 100 SNe Ia to date,
and measurements of their light curves and peak magnitudes find these objects
to be about 0$^m$.25 fainter than predictions for an empty universe. These
measures imply low values for $\Omega_M$ (0.2-0.3) and a positive cosmological
constant, with high statistical significance. Searches out to z$\sim$1-1.2 for
SNe Ia (peak magnitudes m$_I\sim24.5$) will greatly aid in confirming this
result, or demonstrate the existence of systematic errors. Multi-epoch spectra
of SNe Ia at z=0.5 are needed to constrain possible evolutionary
effects. I band searches should be able to find SNe Ia out to z$\sim$2. We
discuss some simulations of deep searches and present histograms of type Ia
and type II SNe discovery statistics at several redshifts.

\end{abstract}

\section{Cosmology with SNe Ia}

Measuring the global curvature and cosmic deceleration (often parameterized as 
$q_0$) of the Universe has been a
fundamental question in astronomy ever since Robertson(1936) and Walker(1936) first
formulated the metric for a homogeneous and isotropic universe. 
Although many methods for measuring the global properties of space exist,
none has yet delivered anything close to a definitive measurement.
By tracing how a standard candle dims as a function of redshift,
usually shown as a Hubble diagram, the effects of global curvature and
deceleration can be seen and quantified.

Two groups are currently using intensive observations of SNe Ia to map the
universe in search of these effects. Our group, the High-Z SN search, is led by Brian Schmidt at
MSSSO. The other effort the Supernovae Cosmology Project, is led by
S. Perlmutter at the Lawrence Berkeley Laboratory.

 Local (z $<$0.15) calibration is provided by $\sim$50 SNe, the
majority from the Calan-Tololo search (Hamuy et al. 1996) and the  work
of the Harvard group
(Riess et al. 1999). Following work by Phillips (1993), it is clear that the
maximum luminosity of a SN Ia event is highly correlated with its decline rate
from maximum light, which can then be
used to improve the accuracy of SNe Ia distances (Hamuy et al 1996b; Riess et
al. 1995).

\begin{figure}[ht]
\vspace*{75mm} 
\includegraphics{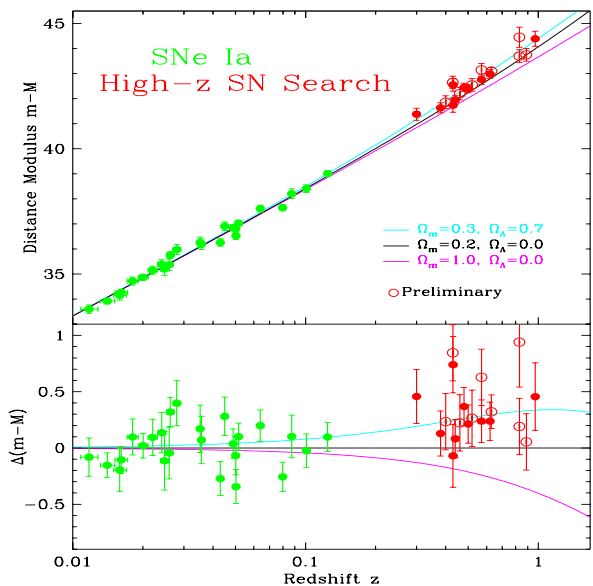}
\caption{The Hubble diagram for Type Ia supernovae, showing luminosity
distance as a function of redshift in the upper panel, and residuals
with respect to an empty universe in the bottom panel.}
\end{figure}

For the high-z sample, both groups primarily search at the prime focus of the
CTIO Blanco 4m, using the BTC, a mosaic CCD camera provided by J. A. Tyson
and G. Bernstein. With 5-minute integrations and a $29^\prime$ field 
we reach a limit of $m_{\rm R} \sim 24$, depending on image quality, 
and can survey approximately 6 square degrees per night (Schmidt et
al. 1998). With deeper I band
searches at CTIO and at the CFHT with the University of Hawaii 8k mosaic
imager, we have found SNe out to redshifts of z$\sim$1. Spectra for redshifts
and classification are taken at ESO, the MMT and Keck, and addition photomtery
for light curve fitting has used 2-4m telescopes at WIYN, ESO, MDM, CTIO and ARC.

Recent work by both teams (Riess et al. 1998, Permutter et al. 1999) has 
shown that Type Ia supernovae at 
z$\sim$0.5 are about $0.^m25$ fainter than the simple predictions for an empty
universe. Figure 1 shows the latest version of the  Hubble diagram from the
High-Z  team. Our value of $\Omega_M$ is low, $\sim$0.24-0.28,
depending analysis techniques and exact sample composition. More surprisingly,
the analysis yield a conclusion that $\Omega_{\Lambda}$  $>$ 0 at a more than
3$\sigma$ confidence level. These measurements, particularly when combined 
with the current measures of the
first Doppler peak of the cosmic background radiation (Hancock et al. 1998),
provide strong evidence for $\Omega_T \sim 1$ (Garnavich et al. 1998; Efstathiou
et al 1999).  Our analysis yields $\Omega_M + \Omega_{\Lambda} =0.94 \pm 0.26$.

\begin{figure}[ht]
\vspace*{55mm} 
\includegraphics{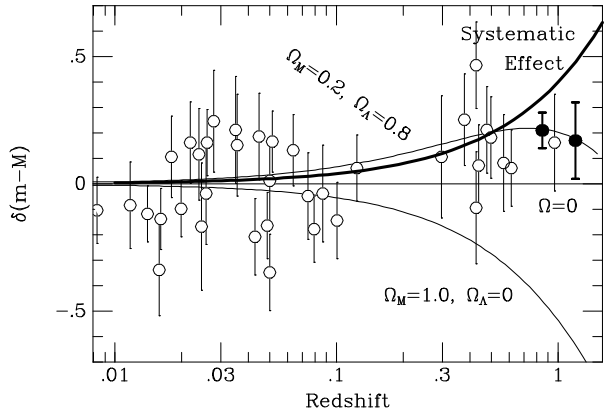}
\caption{The SNIa data are plotted relative to an empty universe.  The
thin curves are the best fitting $\Lambda$ model, and an $\Omega_M=1$
model.  The heavy curve shows an example of a systematic bias which increases
linearly with $z$ and is consistent with our $z=0.5$ data.  The dark
points show the redshifts and experimental uncertainty of the proposed
observations at $z = 0.85$ and $z = 1.2$.}
\end{figure}

Figure 2 shows that the maximum effect of the presence of a positive
cosmological constant (with respect to an empty universe) occurs around
z$\sim$0.85. If the faintness of the SNe Ia at z=0.5 is due to a positive
$\Omega_{\Lambda}$, then at higher redshift supernovae should not get
fainter. By z=1.2, they should even begin to brighten with respect to an empty 
universe. On the other hand, if the fading is due to some systematic error,
such as grey dust or evolution,
one would expect the faintness of to increase with redshift.  In Figure 2 we
show what happens if there is a systematic error that is consistent with our z=0.5
data, but grows linearly with z. Thus one of the major efforts we are
currently undertaking is to search for SNe between z=0.8 and z=1.2.

\subsection{Spectroscopy of Faint SNe Ia}

\begin{figure}[ht]
\vspace*{75mm} 
\includegraphics{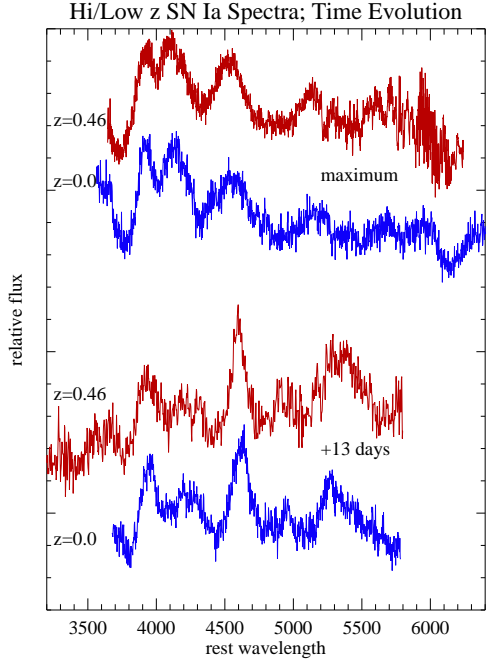}
\caption{Spectral comparison (in f$_\lambda$) of SN 1998Q (z=0.46) with a low
redshift SN Ia at a similar age (t$\sim$0d and 13d).}
\end{figure}

One of the most worrisome systematic effects is source evolution.
Evolution has been the death of other methods used to trace luminosity distances with
redshift. There are strong hints from the low-redshift samples (Hamuy et al. 1996a) 
that the most luminous supernovae are found in later type galaxies, and that
the SN Ia rate per unit luminosity is twice as high in late-type galaxies as in
early-type.  This suggests
that the luminosity of the supernovae in a galaxy depends on the age of the stellar
population.  
Recent attempts (Hoflich, Wheeler \& Thielemann 1998) to model effects of
metallicity, C/O ratio, or changes in  progenitor age are still preliminary. 
The metallicity changes are greatest in
rest-frame UV spectra or light curves (e.g., rise times), and might be discernible in high S/N
spectra. Younger white dwarf progenitors are expected to have a lower C/O
ratio, and this could redden the SN at maximum and steepen the luminosity 
decline.

 If evolution
of SNe Ia is causing the objects to be fainter at z$\sim$0.5 (contrary to
initial model predictions), we would expect this effect to continue at higher
redshifts and thus deeper searches will be helpful as mentioned above.  But
we are also concentrating on obtaining high quality data on our z=0.5
sample. We are obtaining rest frame B and V light curves with HST for 4 SNe in tandem
with multi-epoch Keck spectra in order to compare rigorously the distant
objects to the nearby sample. 

In Figure 3 we show  a spectral comparison for a 
supernova discovered in Jan 1999 (Filippenko et al. 1999). 
To obtain good S/N ($>$10:1) at moderate
resolution (R$\sim$1000) at magnitudes fainter than m$_R$$\sim$23 requires
exposures of several hours with the Keck 10m telescope. 
To obtain similar quality spectra at z$\sim$1
requires working at magnitude m$_I$$\sim$24-25, which will be very challenging
for the current generation of ground based telescopes. Up to this point,
comparisons of spectra at z=0.5  show no significant differences from 
their low redshift brethren (see also Riess et al. 1998).

\section{ Supernovae at z$>$1}

Due to the rapid flux increase above 2600\AA, SNe Ia can be found in I band 
images from CCD cameras out to redshifts of z$\sim$2. While most such efforts 
continue to use ground-based techniques,
because of the wide field capabilities of those telescopes, several SNe have been
discovered using HST. Gilliland, Nugent and Phillips (1999; hereafter GNP)
detected two likely supernovae events in a revisit to the Hubble Deep Field (HDF)
with WF/PC2. With a baseline of 2 years and a shorter I band (only) exposure
(63000 sec), two objects were detected at m$_I$$\sim$26 and 27. These were
associated with galaxies at redshifts 0.95 (spectroscopic) and 1.32
(photometric). This second event is the highest-z SN Ia discovered to date 
(although of course no confirming spectrum of the SN was obtained), and
appears to be associated with an elliptical galaxy, based on colors and 
profile fitting.

GNP performed detailed calculations to quantify the number of expected SNe in
the HDF at a given epoch. The number of SNe expected depends on: 1) the
adopted cosmology; 2) the SN rate per unit redshift; and 3) modeling of the
length of time a given SN remains visible. Calculation of the rates involves
models for the progenitors of types I and II supernovae, assumed initial mass
functions and parameterized delay times  after White
Dwarf formation. These calculations follow studies of the star formation
history of the universe (e.g.,  Madau, Valle \& Panagia 1998). Knowledge of
both the light curves and spectral evolution of SNe is necessary to model the 
length of time  SNe at different redshift remain visible, and we are limited
by our imperfect knowledge of the UV behavior of SNe. The heterogeneous nature
of type II SNe also contributes to the complexity of these calculations.

In Figure 4 (from GNP) is shown the light curves of extinction-free SNe Ia
between 0.25$ < z <$ 1.75. These involve adoption of a zero-point for SN Ia
(e.g., Suntzeff et al. 1999), and a characterization of the light curve shape,
which relates the peak magnitude, color, and decay time of the SN event. GNP chose
to parameterize these effects using the stretch-factor, s, (following Perlmutter et
al. 1999). A stretch-factor of s=1 is used for the calculations in Figure
4. Interpolation of existing spectra for SNe Ia (particularly in the UV), and
a detailed prescription for the K-correction is necessary to perform these
calculations.  A similar analysis is used by GNP to produce light curves for type II SNe.

\begin{figure}[h]
\vspace*{65mm}
\includegraphics{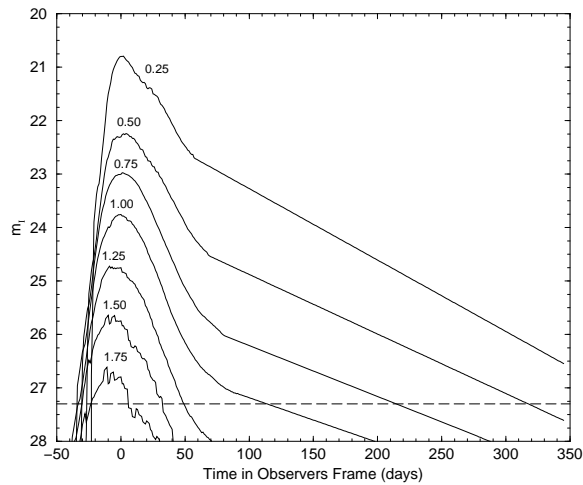}
\caption{ I band light curves for SNe Ia for z=0.25 to z=1.75 for an event of
average intrinsic brightness(from Gilliland et. al).}
\end{figure}

GNP then proceed to calculate rates for the HDF field, based on observed rates
at z=0.5 for SNe Ia (Pain et al. 1996) and Cappellaro et al. (1997) for type
IIs. Figure 5 from GNP shows histograms based on Monte Carlo simulations 
of the relative number of SNe per
magnitude interval in the HDF for redshifts of z=0.95 and 1.3
respectively. For the SNe Ia, they assume a constant volume rate as a
function of redshift, a rather conservative approach. 
For the type II events they use an enhancement in SFR between z=0 and z=2 of
a factor of 4, rather than more extreme values of up to 10 which have
sometimes been proposed.

For the lower redshift case (Figure 5a), the histograms are consistent with
existing ground-based searches. Down to 24th magnitude, type II SNe appear at
10-20\% of the frequency of type Ia. At any given redshift, the SNe Ia events
peak about 1.5-2 magnitudes brighter than the type II events. At fainter
magnitudes at a given redshift, the type II objects will dominate mainly due
to their higher intrinsic rate. The results at z=1.3 are qualitatively similar
in nature, although shifted about 1 mag fainter, and the type II peak at
m$_I$$\sim$27 is compressed due to their cuttoff limit at 27.7.

 Due to the uncertainties of many of the input parameters and the variation of
 cosmic SFR with redshift, the absolute rates predicted in any 
 simulations such as these are uncertain by factors of 2-3. Dahlen and Fransson (1999) predict
a total about 300 SNe per square degree down to I$_{AB}$=27, about 1/3 of
these being  type Ia. This limit is likely reachable for imaging surveys with
8-10m ground-based telescope. A typical redshift will be z$\sim$1, which an
 extended redshift distribution to z$\sim$2. They also predict that NGST will
 be able to find $\sim$50 SNe per pointing (at a limit of K$^{'}$=31.4), about
 20\% of these type Ia.  About 1/3 of the type II SNe should have z$>$2.

\begin{figure}[h]
\vspace*{55mm}
\begin{minipage}[b]{0.46\linewidth}
\includegraphics{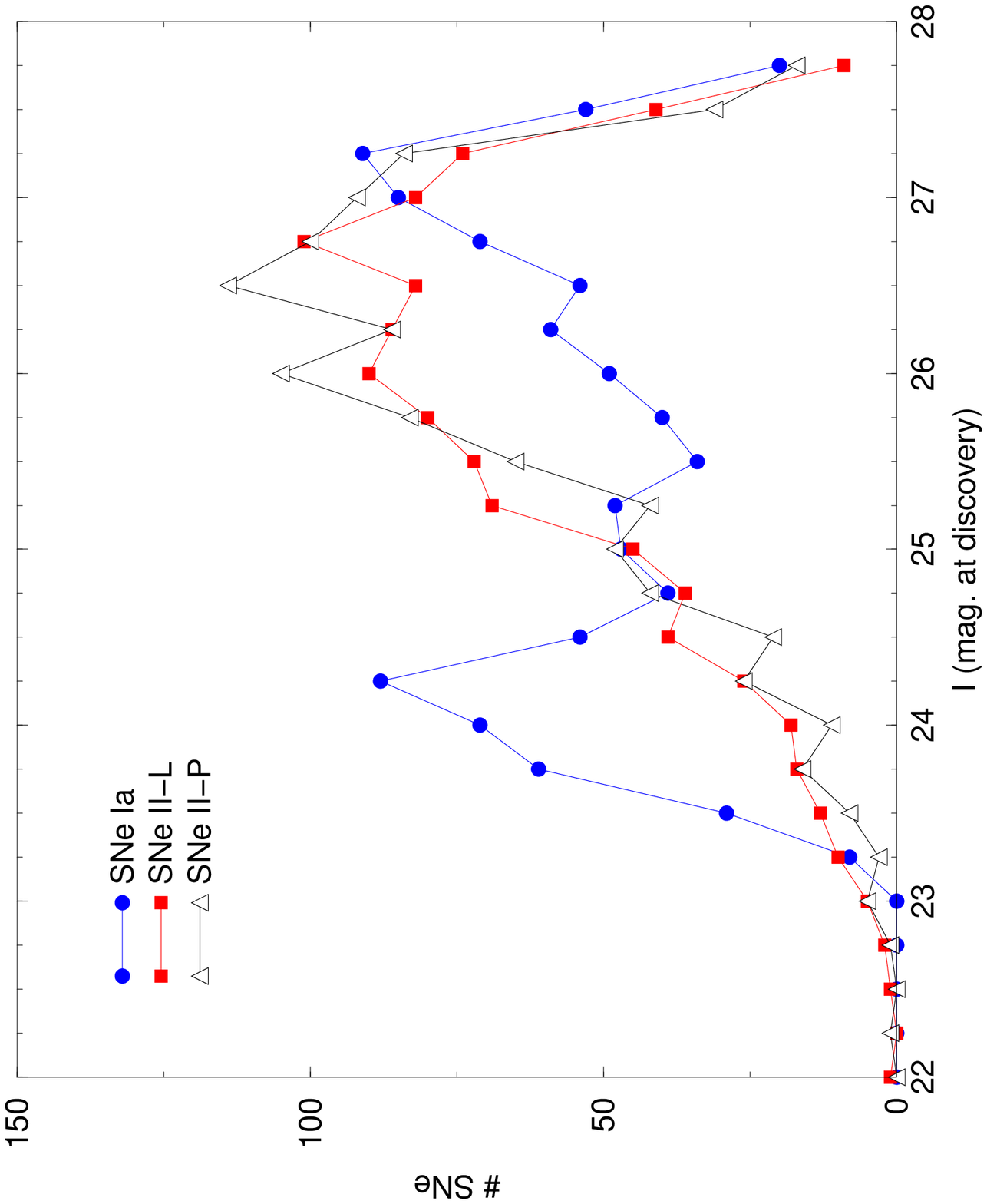}
\centerline{(a)} 
\end{minipage}\hfill
\begin{minipage}[b]{0.46\linewidth}
\includegraphics{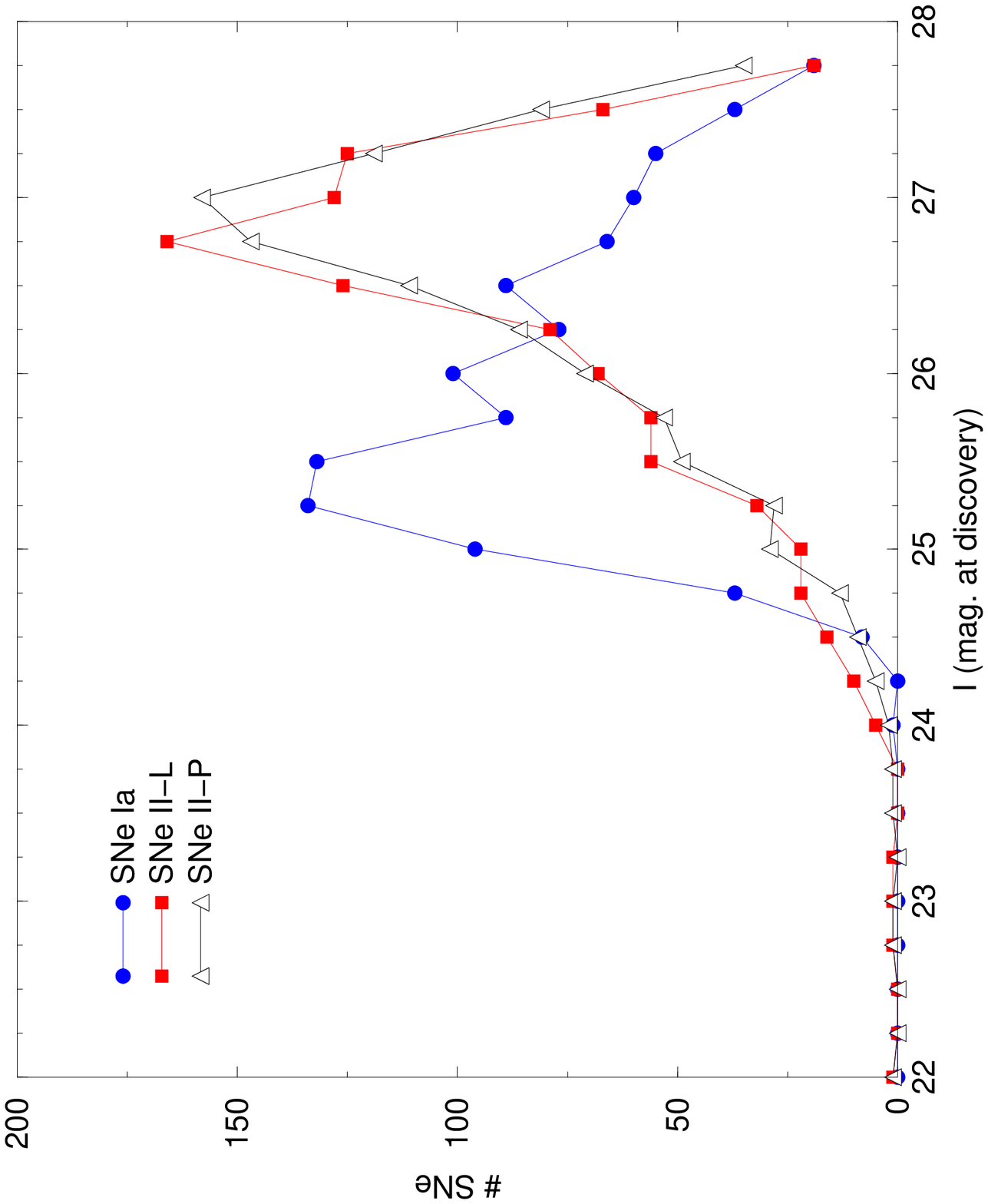}
\centerline{(b)} 
\end{minipage}
\vskip 8pt
\caption{Histogram of (simulated) discovery magnitudes for three types of SNe
at z=0.95 (a) and z=1.3 (b). From Gilliland et. al.}
\end{figure}
  
Sorting out the redshifts and types of these SNe will be a challenging
task. Direct spectroscopy will be very difficult for existing 8-10m telescopes
below 25th magnitude, i.e., a couple of magnitudes brighter than the imaging
limit. In principle the SNe type can be determined from careful
photometry of the light curves. In order to perform such a characterization,
however, the decline must be followed with multiple observations (7-10) over more
than a month, while the SN fades 1-2 magnitudes. Thus light curves are also
limited to objects 1-2 magnitudes above the discovery limit. Photometric 
redshifts of the host galaxies may be possible in
some cases, although a significant fraction of the SNe occur in dwarf or low
surface brightness objects; in existing 4m SNe searches the SNe is almost
always significantly brighter than its host. Dahlen and Fransson (1999)
discuss the possibility of determining photometric redshifts from the SN light
directly, but the time evolution and variety of spectral energy distributions
for the range of SNe types makes this a very difficult proposition. Thus it is
hard to escape the conclusion that we soon will need spectroscopic capability
at m$_I$$\sim$26-27, and this need will become even more important with the
arrival of capabilities like NGST.

\section{Other Things that Erupt in the Night}
 During our high-z searches we have discovered a few objects which fade very
 rapidly. We have at least 4 of these objects, approximately  a factor of 10
 down from our SNe Ia detection rate. Peak magnitudes are m$_R$$\sim$23-24,  with
a time scale of between a few hours and a few days. These objects
 are hard to detail, since so little follow up observations are available.
One possibility is that these objects are ``optical'' GRBs, i.e., we are
 detecting unbeamed optical radiation from a GRB pointed in some other
 direction. Another possibility is that these are SNe II in the brief high
 UV/optical luminosity phase, immediately after shock breakout (Blinnikov et al. 1998).

\acknowledgements
I would like to thank the members of my team for providing much of the
information for this talk and paper, particularly Brian Schmidt, Adam Riess,
Alex Filippenko and Mario Hamuy. Special thanks to Peter Nugent for providing
versions of his calculation. The CTIO, ESO, CFHT, Keck and HST TACs and directors have also been
generous in their time allocations, which has been vital to the assemblage of
the detailed data necessary for a project of this scope.


\end{document}